\documentclass[sort&compress,final]{aipproc}
\layoutstyle{8x11double}

\def\Journal#1#2#3#4{{\it #1} {\bf #2}, #3 (#4)}

\def\PLB{{\em Phys. Lett.}  B}
\def\PRL{\em Phys. Rev. Lett.}
\def\PRD{{\em Phys. Rev.} D}

\def\be{\begin{equation}}
\def\ee{\end{equation}}
\def\bea{\begin{eqnarray}}
\def\eea{\end{eqnarray}}

\begin{document}
\title{Three Tests of LambdaCDM}


\author{C. J. A. P. Martins}{
  address={Centro de Astrof\'{\i}sica, Universidade do Porto, Rua das Estrelas, 4150-762 Porto, Portugal}
}

\begin{abstract}
The observational evidence for the acceleration of the universe demonstrates that canonical theories of gravitation and particle physics are incomplete, if not incorrect. A new generation of astronomical facilities will shortly be able to carry out precision consistency tests of the standard cosmological model and search for evidence of new physics beyond it. I describe some of these tests, focusing on the universality of nature's fundamental couplings and the characterization of the properties of dark energy. I will also comment on prospects for forthcoming ESA and ESO facilities in which the CAUP Dark Side team is involved.
\end{abstract}

\maketitle


\section{The dark side of the universe}

In the middle of the XIX century Urbain Le Verrier and others mathematically discovered two new planets by insisting that the observed orbits of Uranus and Mercury agreed with the predictions of Newtonian physics. The first of these--- which we now call Neptune---was soon observed by Johann Galle and Heinrich d'Arrest. However, the second (dubbed Vulcan) was never found. We now know that the discrepancies in Mercury's orbit were a consequence of the fact that Newtonian physics can't adequately describe Mercury's orbit, and accounting for them was the first success of Einstein's General Relativity.

Over the past several decades, cosmologists have mathematically discovered two new components of the universe---which we have called dark matter and dark energy---which have so far not been directly detected. Whether they will prove to be Neptunes or Vulcans remains to be seen but even their mathematical discovery highlights the fact that the standard $\Lambda$CDM paradigm, despite its phenomenological success, is at least incomplete.

Something similar applies to particle physics, where to some extent it is our confidence in the standard model that leads us to the expectation that there must be new physics beyond it. Neutrino masses, dark matter and the size of the baryon asymmetry of the universe all require new physics, and---significantly---all have obvious astrophysical and cosmological implications.

Recent years have made it clear that further progress in fundamental particle physics will increasingly depend on progress in cosmology. One must therefore carry out explicit consistency tests of the standard cosmological model and search for evidence of new physics beyond it.

\section{Fundamental scalar fields}

After a quest of several decades, the recent LHC evidence for a Higgs-like particle \cite{ATLAS,CMS} finally provides strong evidence in favour of the notion that fundamental scalar fields are part of Nature's building blocks. A pressing follow-up question is whether the associated field has a cosmological role (or indeed if there is some cosmological counterpart).

At the same time, fundamental scalar fields already play a key role in most paradigms of modern cosmology. {\it Inter alia} they are routinely invoked to describe
\begin{itemize}
\item A period of exponential expansion of the early universe (inflation).
\item Cosmological phase transitions and their relics (cosmic defects).
\item The dynamical dark energy which may be powering the current acceleration phase.
\item The possible spacetime variation of nature's fundamental couplings.
\end{itemize}

Even more important than each of these paradigms is the fact that they don't occur alone: whenever a scalar field plays one of the above roles, it will also leave imprints in other contexts that one can look for. Three obvious examples can be given
\begin{itemize}
\item In realistic models of inflation, the inflationary phase ends with a phase transition at which cosmic defects will form; the energy scales of both will therefore be unavoidably related.
\item A particular type of (so-called 'frustrated') cosmic defect networks has been invoked as a possible explanation for dark energy; however this possibility is now excluded, both on observational grounds and because the detailed dynamical properties required do not in fact occur for realistic defect networks.
\item In realistic models of dark energy, where the dark energy is due to a dynamical scalar field, this field will couple to the rest of the model and lead to potentially observable variations of nature's fundamental couplings; we will return to this point later in this contribution.
\end{itemize}
Although this complementary point is often overlooked, it will be crucial for future consistency tests.

\section{Varying fundamental couplings}

Nature is characterized by a set of physical laws and fundamental dimensionless couplings, which historically we have assumed to be spacetime-invariant. For the former this is a cornerstone of the scientific method (indeed, it's hard to imagine how one could do science at all if it were not the case), but for the latter it is only a simplifying assumption without further justification. 
These couplings determine the properties of atoms, cells, planets and the universe as a whole, so it's remarkable how little we know about them---in fact we have no 'theory of constants', that describes their role in physical theories or even which of them are really fundamental. If they vary, all the physics we know is incomplete.

Fundamental couplings are expected to vary in many extensions of the current standard model. In particular, this will be the case in theories with additional spacetime dimensions, such as string theory. Interestingly, the first generation of string theorists had the hope that the theory would ultimately predict a unique set of laws and couplings for low-energy physics. However, following the discovery of the evidence for the acceleration of the universe this hope has been pragmatically replaced by an 'anything goes' approach, sometimes combined with anthropic arguments. Regardless of the merit of such approaches, experimental and observational tests of the stability of these couplings may be their best route towards a testable prediction.

It goes without saying that a detection of varying fundamental couplings will be revolutionary: it will immediately prove that the Einstein Equivalence Principle is violated (and therefore that gravity can't be purely geometry), that there is a fifth force of nature, and so on.

Moreover, even improved null results are important and useful. The simple way to understand this is to realize that the natural scale for cosmological evolution of one of these couplings (driven by a fundamental scalar field) would be Hubble time. We would therefore expect a drift rate of the order of $10^{-10}$ yr${}^{-1}$. However, current local bounds, coming from atomic clock comparison experiments, are 6 orders of magnitude stronger~\cite{Rosenband,Ferreira}.

Recent astrophysical evidence from quasar absorption systems~\cite{Webb} observed with HIRES/Keck and UVES/VLT suggests a parts-per-million spatial variation of the fine-structure constant $\alpha$ at low redshifts; although no known model can explain such a result without considerable fine-tuning, it should also be said that there is also no identified systematic effect that can explain it \cite{Thesis}.

One possible cause for concern (with these and other results) is that almost all of the existing data has been taken with other purposes in mind (and subsequently re-analized for this purpose), whereas this kind of measurements needs customized analysis pipelines and wavelength calibration procedures beyond those supplied by standard pipelines~\cite{Rodger}.

An ongoing ESO UVES Large Programme dedicated to fundamental physics will soon provide further measurements \cite{Molaro}. This is the only large program (so far) dedicated to varying constants, carrying out observations of a selected sample with optimized methodology. Although the programme was selected before the $\alpha$ dipole was known (and it is therefore not optimized for it), it will be able to test it.

In the short term the PEPSI spectrograph at the LBT can also play a role here, and in the longer term a new generation of high-resolution, ultra-stable spectrographs like ESPRESSO (for the VLT) and CODEX (or its future incarnation, for the E-ELT) will significantly improve the precision of these measurements and should be able to resolve the current controversy. A key technical improvement will be that ultimately one must do the wavelength calibration with laser frequency combs.

In theories where a dynamical scalar field yields varying $\alpha$, the other gauge and Yukawa couplings are also expected to vary. In particular, in Grand Unified Theories the variation of $\alpha$ is related to that of energy scale of Quantum Chromodynamics, whence the nucleon masses necessarily vary when measured in an energy scale that is independent of QCD (such as the electron mass).

It follows that we should expect a varying proton-to-electron mass ratio, $\mu=m_p/m_e$, which can be probed with molecular Hydrogen \cite{Thompson} and other molecules. These use the fact that molecular vibrational and rotational transitions depend on the reduced mass of the molecule, and the dependence is different for different transitions. Obviously, the specific relation between $\alpha(z)$ and $\mu(z)$ will be completely model-dependent, but this is a blessing rather than a curse: astrophysical measurements of both provide us with a unique discriminating tool between different unification models.

It's worth emphasizing that while molecular Hydrogen measurements do probe $\mu$, those involving more complicated molecules are probing an effective nucleon-to-electron mass ratio, and this will be proportional to $\mu$ (the proportionality factor being a pure number) only in the absence of composition-dependent forces. But again, this is provides us with a unique opportunity: by simultaneously doing these measurements with several molecules ($H_2$, $HD$, ammonia, etc), which occasionally may be found in the same cloud, one will ultimately be able to constrain possible composition-dependent couplings. This is a direct astrophysical test of the Equivalence Principle, which is not feasible with current facilities but should be within the reach of ESPRESSO and CODEX.

At much higher redshifts, the Cosmic Microwave Background is an ideal, clean probe of a varying fine-structure constant. A changed $\alpha$ will impact the ionization history of the universe: the energy levels and binding energies are shifted, and the Thomson cross-section is proportional to $\alpha^2$. Having said this, bounds are relatively weak due to degeneracies with other cosmological parameters, and the percent barrier has only recently been broken~\cite{Eloisa09}.

For the reasons explained above it is too naive to consider variations of a single quantity, and fortunately the CMB data is becoming good enough for coupled variations to be constrained, despite the degeneracies; for example \cite{Martins10} constrained simultaneous variations in $\alpha$ and the gravitational sector. The latter can be interpreted as constraints on Newton's gravitational constant $G$ if one chooses units in which the particle masses are fixed.

A cosmological constant is negligible at recombination, but a dynamical, tracking scalar field can induce significant $\alpha$ variations. An example is the class of early dark energy models \cite{Doran}, linearly coupled to electromagnetism \cite{Nunes}. One can in fact constrain the coupling between the putative scalar field and electromagnetism, independently (and on a completely different scale) from what is done in local tests~\cite{Calabrese}. The local bound is (conservatively) at the $10^{-3}$ level, and our constraint is only about 20 times weaker, which is a testimony to the CMB sensitivity. (As a comparison, lensing constraints on the Eddington parameter $\gamma$ \cite{Schwab} are currently 2500 times weaker than those from the Cassini bound \cite{Bertotti}.)

The recent CMB measurements from WMAP and arcminute angular scales (from ACT and SPT) suggest that the effective number of relativistic degrees of freedom is larger than the standard value of
\be
N_{\rm eff} = 3.04\,,
\ee
and inconsistent with it at more than  two standard deviations. We have recently shown~\cite{Eloisa12} that, if one assumes this standard value, these same CMB datasets significantly improve previous constraints on ${\alpha}$, with
\be
\frac{\alpha}{\alpha_0} = 0.984 \pm 0.005\,,
\ee
hinting also to a more than two-sigma deviation from the current, local, value. A significant degeneracy is present between ${\alpha}$ and $N_{\rm eff}$, and when variations in the latter are allowed the constraints on ${\alpha}$ are consistent with the standard value. Again it's worth stressing that deviations of either parameter from their standard values would imply the presence of new, currently unknown physics. Forthcoming Planck data should improve these constraints. Moreover, once this data is available, and additional set of Equivalence Principle tests will also become possible.

Although QSO spectroscopy and the CMB are by now the two standard methods to probe the stability of fundamental couplings, they are merely the tip of the iceberg. Many compact astrophysical objects can also be used for this purpose, and in particular to test for their possible environmental dependence. Recent progress includes work on
\begin{itemize}
\item Population III stars~\cite{Ekstrom},
\item Neutron stars~\cite{Angeles}, and
\item Solar-type stars~\cite{Vieira}.
\end{itemize}
Naturally all these systems are sensitive to several dimensionless couplings (and not just $\alpha$) so they will soon provide us with further opportunities to constrain unification scenarios.

\section{Dynamical dark energy}

Observations suggest that the universe is dominated by an energy component whose gravitational behavior is quite similar to that of a cosmological constant. Its value is so small that a dynamical scalar field is arguably a more likely explanation. Such a field must be slow-rolling (which is mandatory for $p<0$) and be dominating the dynamics around the present day. It follows~\cite{Carroll} that couplings of this field to the rest of the model lead to potentially observable long-range forces and time dependencies of the constants of nature.

The above point is worth emphasizing because it is often misunderstood. Any scalar field couples to gravity. It couples to nothing else only if there is an {\it ad-hoc} global symmetry which suppresses couplings to the rest of the Lagrangian. (In that case, only derivatives and derivative couplings will survive.) However, such symmetries are hard to come by. Specifically, quantum gravity effects don't respect global symmetries, and there are no unbroken global symmetries in string theory. Therefore, if one goes to great lenghts to justify a coupling one is in fact reversing the burden of proof: the expectation is that scalars in the theory will couple to the rest of the world in any manner not prevented by symmetry principles. (It is the absence of couplings that requires proper justification.)

Standard observables such as supernovae are of limited use as dark energy probes~\cite{Maor}. A clear detection of varying $w(z)$ is crucial, given that we know that $w\sim-1$ today. Since the field is slow-rolling when dynamically important (close to the present day), a convincing detection of a varying $w(z)$ will be tough at low redshift, and we must probe the deep matter era regime, where the dynamics of the hypothetical scalar field is fastest. Varying fundamental couplings are ideal for probing scalar field dynamics beyond the domination regime~\cite{Nunes}: such measurements can presently be made up to redshift $z\sim4$, and future facilities such as the E-ELT may be able to significantly extend this redshift range. Importantly, even null measurements of varying couplings can lead to interesting constraints on dark energy scenarios \cite{Coupling1,Coupling2}.

We have recently studied~\cite{Amendola}, using Principal Component Analysis techniques, the impact of ESPRESSO and CODEX in constraining dark energy through measurements of varying fundamental couplings. In the case of CODEX, a reconstruction using quasar absorption lines is expected to be more accurate than using supernovae data (its key advantage being huge redshift lever arm), and even ESPRESSO can provide a significant contribution. Since the two types of measurements probe different (but overlapping) redshift ranges, combining them leads to a more complete picture of the evolution of the equation of state parameter, and these can realize the prospect of a detailed characterization of dark energy properties almost all the way up to redshift 4.

Although the most obvious way to proceed is to combine the two datasets, we should also point out that they can be used separately to provide independent reconstructions. Comparing the two reconstructions will in itself provide a consistency test, specifically for the assumption on the coupling between the scalar field and electromagnetism. In this case one can also obtain a measurement for the coupling parameter, which can be compared to that obtained from the CMB (as discussed above) and those obtained from local tests. Again, null results can also be constraining---see \cite{Coupling1,Coupling2} for tight constraints on a broad range of slow-roll models.

Dark energy reconstruction using varying fundamental constants does in principle require a mild assumption on the field coupling, but there are in-built consistency checks, so that inconsistent assumptions can be identified and corrected. An explicit example of an incorrect assumption that leads to an observation inconsistency is discussed in \cite{Pauline}. In this regard CODEX at the E-ELT, with its ability to carry out the Sandage-Loeb test \cite{Sandage,Loeb}, will play a crucial role \cite{OurSL1}. Interesting synegies also exist between these ground-based spectroscopic methods and Euclid, which need to be further explored.

\section{The quest for redundancy}

Whichever way one finds direct evidence for new physics, it will only be believed once it is seen through multiple independent probes. This was manifest in the case of the discovery of the recent acceleration of the universe, where the supernova results were only accepted by the wider community once they were confimed through CMB, large-scale structure and other data.

It is clear that history will repeat itself in the case of varying fundamental couplings and/or dynamical dark energy. It is therefore crucial to develop consistency tests---in other words, astrophysical observables whose behaviour will also be non-standard as a consequence of either or both of the above.

The temperature-redshift relation,
\be
T(z)=T_0(1+z)\,,
\ee
is a robust prediction of standard cosmology; it assumes adiabatic expansion and photon number conservation, but it is violated in many scenarios, including string theory inspired ones. At a phenomenological level one can parametrize deviations to this law by adding an extra parameter, say
\be
T(z)=T_0(1+z)^{1-\beta}\,.
\ee
Measurements of the SZ effect at resdshifts $z<1$, combined with spectroscopic measurements at redshifts $z\sim2-3$ yield the direct constraint \cite{Noterdaeme}
\be
\beta=-0.01\pm0.03\,.
\ee
Our recent work \cite{Atrio,Gemma} has shown that forthciming data from Planck, ESPRESSO and CODEX will lead to much stronger constraints:
\begin{itemize}
\item Planck on its own can be as constraining as the existing bound,
\item ESPRESSO can improve on the current constraint by a factor of about 3, and
\item CODEX will improve on the current bound by one order or magnitude.
\end{itemize}
We emphasize that estimates of all these gains rely on quite conservative on the number of sources (SZ clusters and absorption systems, respectively) where these measurements can be made. If the number of such sources increases, future constraints can be correspondingly stronger.

The distance duality relation,
\be
d_L=(1+z)^2d_A\,,
\ee
is an equally robust prediction of standard cosmology; it assumes a metric theory of gravity and photon number conservation, but is violated if there's photon dimming, absorption or conversion. At a similarly phenomenological level one can parametrize deviations to this law by adding an extra parameter, say
\be
d_L=(1+z)^{2+\epsilon}d_A\,.
\ee
In this case, current constraints are \cite{Avgoustidis}
\be
\epsilon=-0.04\pm0.08\,,
\ee
and improvements are similarly expected from Euclid, the E-ELT and JWST.

In fact, in many models where photon number is not conserved the temperature-redshift relation and the distance duality relation are not independent. With the above parametrizations it's easy to show~\cite{Gemma} that
\be
\beta=-\frac{2}{3}\epsilon\,,
\ee
but one can in fact further show that a direct relation exists for any such model, provided the dependence is in redshift only (models where there are frequency-dependent effects are more complex). This link allows us~\cite{Gemma} to use distance duality measurements to further constrain $\beta$, and we recently found
\be
\beta=0.004\pm0.016\,
\ee
up to a redshift $z\sim 3$, which is a $40\%$ improvement on the previous constraint. With the next generation of space and ground-based experiments, these constraints can be further improved (as discussed above) by more than one order of magnitude. 

\section{Outlook}

The observational evidence for the acceleration of the universe demonstrates that canonical theories of cosmology and particle physics are incomplete, if not incorrect. Several few-sigma hints of new physics are emergine, but so far these are smoke without a smoking gun; it's time to actively search for the gun.

The forthcoming generation of high-resolution ultra-stable spectrographs will play a key role in this endeavour, by enabling a new generation of precision consistency tests of the standard cosmological paradigm and its extensions. Some further exciting  possibiblites, only mentioned briefly in this contribution, include direct astrophysical Equivalence Principle tests and E-ELT measurements of the redshift drift.

Last but not least, there are important synergies between ground and space facilities, and in  particular between the E-ELT andEuclid, which require further study: together, they will be a unique tool to study fundamental physics and gravity.


\begin{theacknowledgments}
I am grateful to Mariusz Dabrowski and the rest of the Multicosmofun'12 organizers for their hospitality and for organizing such an enjoyable and productive meeting.

This work has been done in the context of the grant PTDC/FIS/111725/2009 (The Dark Side of the Universe, funded by FCT), with additional support from grants PESSOA 2012/2013 Proc. 441.00 (Testing Fundamental Physics with Planck, finded in Portugal by FCT) and PP-IJUP2011-212 (Astrophysical Tests of Fundamental Physics, funded by U. Porto and Santander Totta).

Many interesting discussions with the rest of CAUP's Dark Side team and our collaborators elsewhere have shaped my views on this subject, and are gratefully acknowledged. The work of CJM is supported by a Ci\^encia2007 Research Contract, funded by FCT/MCTES (Portugal) and POPH/FSE (EC).
\end{theacknowledgments}

\bibliographystyle{aipproc}   

\end{document}